\documentclass[a4paper,fleqn]{cas-sc}

\usepackage[authoryear]{natbib}
\usepackage{caption}
\usepackage{subcaption}
\usepackage[utf8]{inputenc}

\def\tsc#1{\csdef{#1}{\textsc{\lowercase{#1}}\xspace}}
\tsc{WGM}
\tsc{QE}
\tsc{EP}
\tsc{PMS}
\tsc{BEC}
\tsc{DE}

\begin{document}
\let\WriteBookmarks\relax
\def\floatpagepagefraction{1}
\def\textpagefraction{.001}

\shorttitle{An ensemble-based framework for mispronunciation detection of Arabic phonemes}

\shortauthors{Selim et~al.}

\title [mode = title]{An ensemble-based framework for mispronunciation detection of Arabic phonemes}                      




\author[a]{Sükrü Selim Calık}
\ead{sselimcalik1@gmail.com}
\author[b]{Ayhan Kucukmanisa}
\ead{ayhan.kucukmanisa@kocaeli.edu.tr}
\author[c]{Zeynep Hilal Kilimci\corref{mycorrespondingauthor}}
\ead{zeynep.kilimci@kocaeli.edu.tr}
\cortext[mycorrespondingauthor]{Corresponding author: Zeynep Hilal Kilimci}

\address[a]{Maviay Consultancy Company, Kocaeli University Technopark, 41275, Kocaeli, Turkey}
\address[b]{Department of Electronics and Communication Engineering, Kocaeli University, 41001, Kocaeli, Turkey}
\address[c]{Department of Information Systems Engineering, Kocaeli University, 41001, Kocaeli, Turkey}
\begin{abstract}
Determination of mispronunciations and ensuring feedback to users are maintained by computer-assisted language learning (CALL) systems. In this work, we introduce an ensemble model that defines the mispronunciation of Arabic phonemes and assists learning of Arabic, effectively. To the best of our knowledge, this is the very first attempt to determine the mispronunciations of Arabic phonemes employing ensemble learning techniques and conventional machine learning models, comprehensively. In order to observe the effect of feature extraction techniques, mel-frequency cepstrum coefficients (MFCC), and Mel spectrogram are blended with each learning algorithm. To show the success of proposed model, 29 letters in the Arabic phonemes, 8 of which are hafiz, are voiced by a total of 11 different person. The amount of data set has been enhanced employing the methods of adding noise, time shifting, time stretching, pitch shifting. Extensive experiment results demonstrate that the utilization of voting classifier as an ensemble algorithm with Mel spectrogram feature extraction technique exhibits remarkable classification result with 95.9{\%} of accuracy. 

\end{abstract}




\maketitle

\section{Introduction}

Computer Aided Language Learning (CALL) systems are popular nowadays because they facilitate people to progress the ability of language and pronunciation. CALL mainly focus on fields such as pronunciation errors in non-native learners’ talk. CALL achieves tasks, such as detection of pronunciation error, speech recognition, and grading of pronunciation. Furthermore, there are many researches on speech processing, applied in various languages for the purpose of aiding language learning \citep{cucchiarini1998assessment}, \citep{neumeyer2000automatic}, \citet{minematsu2004pronunciation}. Advances in computer science, especially in artificial intelligence, have consented a great deal of research to be done with CALL \citep{cucchiarini1998assessment}, \citep{neumeyer2000automatic}, \citet{minematsu2004pronunciation}, \citep{adnan2012computer}, \citet{hu2013new}, \citep{ehsani1998speech}, \citep{arias2010automatic}.

For a certain language, speakers of distinct languages incline to compose pronunciation errors in their utterances because the muscles of mouth are not able to pronunciate the nuances of the specific language. For this reason, researchers generally focus on to investigate the mispronunciation for various languages such as English, Dutch, and French, however; literature studies on Arabic are limited. Nevertheless, studies on Arabic have been increasing in recent years \citep{abufanas2013computer}, \citet{khan2013automatic}, \citep{muaad2021arcar}, \citep{shaalan2005intelligent}, \citep{nazir2019mispronunciation}, \citep{ziafat2021correct}. Arabic, the most widely spoken \citep{lane201910} with over 290 million native speakers and 132 million non-native speakers, and one of the six official languages of the United Nations (UN), occurs two main dialects, Classical Arabic (CA) and modern standard Arabic (MSA). Classical Arabic is the language of the Quran while MSA is the modified version of the Quran used in daily communication. The pronunciation rules in the language of the Quran are very well defined in order to preserve the correct meaning of the words. This work concentrates on detecting the mispronunciation of the phonemes in the Quran language employing both machine and ensemble learning models. 

Ensemble learning is a technique that employs multiple classifiers to acquire a better success compared to an individual classifier. In other words, a bunch of single classifiers is consolidated for the purpose of composing an ensemble-based classifier. The generation and consolidation steps are the main phases of ensemble learning strategy. In generation phase of ensemble, a diversified data set of single learners is composed of the training set. The final decision is provided by consolidating each individual classifiers in the integration phase. The major approach in ensemble strategy is whence to produce many learners and combine decisions of each individual classifier such that the consolidation of base learners boosts the success of a single classifier \citep{rokach2010ensemble}, \citep{polikar2006ensemble}, \citep{gopika2014analysis}, \citep{ren2016ensemble}, \citet{kilimci2016effectiveness}.

In this study, we concentrate on the detection of mispronunciation of Arabic phonemes observing the effects of both conventional machine learning algorithms and ensemble learning techniques when blended with the feature extraction methods. For this purpose, support vector machine, k-nearest neighbors, decision tree, naive Bayes, random forest classifiers are evaluated diversifying with feature extraction techniques, namely mel-frequency cepstrum coefficients, and Mel spectrogram. After that, each decision of individual classifiers is consolidated majority voting, boosting, bagging, stacking with logistic regression, stacking with random forest, stacking with extra tree methods. Experiment results reveal that the usage of majority voting as an ensemble technique with Mel spectrogram feature extraction method gives considerable results with 95.3{\%} of accuracy. 

The main contributions of the paper are as follows:

$\bullet$ In order to accelerate future studies, our own dataset of Arabic letters is created with hafizes.

$\bullet$ Providing better accuracy with low computational load compared to high-performance deep learning models.

$\bullet$ Suitable for low capacity embedded platforms with its low complexity.

The rest of paper is designed as follows: Section \ref{sec2} introduces related literature studies on mispronunciation of Arabic phonemes. Section \ref{sec3} presents individual machine learning classifiers and ensemble learning methods utilized in this work. Section \ref{sec4} gives experimental results. The paper is concluded with discussion and conclusion in Section \ref{sec5}.

\section{Literature Review}\label{sec2}

In this section, literature studies on mispronunciation of Arabic phonemes are briefly introduced.

\citet{muhammad2010voice} present a non-real-time e-Hafiz system that proposes to avoid the Quran from being mispronunciation. For this purpose, data preparation, feature extraction, and modeling stages are performed. After silence removal and pre-emphasis methods are implemented at the data preparation step, framing, windowing, discrete Fourier transform, Mel Filter-bank, logarithm, and inverse discrete Fourier transformation techniques are applied in the feature extraction phase, respectively. After that, similarity is calculated to find mispronunciation. The authors compare the introduced model with a counterpart web application on five people and inform that they exhibit advancement in their memorization. In other study, \citep{elsayed2019evaluation} aim an automated model to assess the memorization of the Qur’an depending on Hafiz reading. Mel-frequency cepstral coefficient is assessed as feature extraction technique while vector quantization is employed for the purpose of dimension reduction by the proposed system. The authors present that the experiment results ensure remarkable accuracy scores to evaluate Quran memorization with the utilization of proposed system. In another study \citep{putra2012developing}, a speech recognition model is developed to learn the Holy Quran. Gaussian mixture model is blended with MFCC feature extraction technique. Different speech signal processing techniques are applied namely, minimal sampling frequency, frame blocking, windowing, discrete Fourier transform, and dynamical 40 cepstral coefficient and spectrum energy in addition to MFCC. The authors present the experimental results provide 70{\%} of accuracy performance for pronunciation, 90{\%} of accuracy success for reading rules of tajweed, and 60{\%} of accuracy score for the consolidation of both pronunciation and tajweed reading rules.

\citep{arafa2018dataset} introduce a speech recognition system that is able to discover mispronunciation. The dataset is composed of 89 students, which of 46 is female. 28 Arabic phonemes are voiced 10 times. After gathering 890 utterances, MFCC coefficients are extracted for modeling with five different machine learning models. These are k-nearest neighbor, support vector machine, naive Bayes, multi layer perceptron, and random forest. The experimental results indicate that the random forest technique exhibits 85.02{\%} of accuracy result, which is better than compared to other machine learning models. In another recent study, \citep{nazir2019mispronunciation} present a model depending on mispronunciation detection employing deep convolutional neural network properties for Arabic phonemes. As a first, features are extracted various layers of CNN ranges from 4 to 7. After that, k-nearest neighbor, support vector machine, and neural network classifier are employed for training phase. They compare the experiment result when hand-crafted features are employed instead of utilizing deep features. Finally, the authors emphasize that the proposed learning method considerably boosts the success of the mispronunciation determination process in terms of accuracy by 92.2{\%} when deep features are employed. In a study, \citep{akhtar2020improving} introduce a model to develop the detection of mispronunciation of Arabic words for non-native students utilizing features of deep convolutional neural network. To utilize deep features, features are extracted from layers 6, 7, and 8 of Alex Net. After that, n-nearest neighbor, support vector machine, and random forest are employed at the training phase. To show the effect of deep features, authors also evaluate the features extracted using mel frequency cepstral coefficients. The same three model are employed and the results of both approach is compared. They emphasize that the introduced model with 93.2{\%} of accuracy achieves the determination of incorrect pronunciation of Arabic words. \citep{maqsood2016complete} concentrate on the detection of mispronunciation for Arabic phonemes employing acoustic phonetic features (APF) with support vector classifier. For this purpose, authors concentrate on the acoustic phonetic features instead of utilizing confidence measure-based scores. The data set is constructed with 500 utterances of 100 speakers. The features are composed of root mean square energy (RMSE), MFCCs,low energy, spectral features, zero-cross rate, pitch and statistical features. After that, SVM is employed to determine the mispronunciation of Arabic phonemes. The paper is concluded that support vector machine demonstrates 97.5{\%} of accuracy result.

In another study, \citep{farooq2021mispronunciation} focus on determination of mispronunciation in articulation points for Arabic letters. For this purpose, Relative Spectral Transform - Perceptual Linear Prediction feature extraction technique is blended with Hidden Markov Model (HMM). 1,486 samples are gathered for 29 Arabic letters. Various techniques are employed to perform with HMM, like for discovering the observation probability Gaussian mixture models or forward method is utilized. Authors conclude the study that proposed model is capable to recognize the mispronunciation performing 85{\%} of recognition rate. In the study \citep{nazir2021arabic}, support vector machine-based system is investigated for the aim of determining mispronunciation of Arabic phonemes for Asian speakers on 28 Arabic phonemes. The proposed system is accomplished 88{\%} of accuracy. \citep{asif2021approach} concentrate on the correct pronunciation of Arabic vowels with the help of deep neural networks. The new data set is constructed and augmented with various techniques due to the limited number of samples. The data set is composed of 85 individual records which of 43 is female and only four of them are non-native speakers. After gathering 6,229 records, preprocessing stage is carried out. For this purpose, noise reduction, data segmentation, re-sampling, and silence truncation are implemented. To determine the mispronunciation of Arabic vowels, optimized convolutional neural network is modeled. The proposed model shows 95.77{\%} of accuracy score on pronunciation classification of classical Arabic phonemes. \citep{maqsood2017comparative} perform comparative analysis of different classifiers for the purpose of detecting mispronunciation for Arabic phonemes using acoustic phonetic features namely, MFCC. The following machine learning algorithms are employed: Random forest, naive Bayes, Ada-boost, and k-nn. Authors report that random forest model outperforms others with 95.4{\%} of accuracy.

\citep{shareef2022comparison} concentrate on the detailed comparison of feature extraction models for the purpose of detecting impairment Arabic speech. The feature extraction technique is based on diverse versions of wavelet transformation. To detect the impairment Arabic speech, LSTM and CNN-LSTM models are constructed. The combination of MFCC and LSTM achives the best classification accuracy with 93{\%} while it is followed by CNN-LSTM with 91{\%} of accuracy. Mispronunciation detection system is designed by \citep{algabri2022mispronunciation} for non-native Arabic speakers for the purpose of providing them impairment feedback. The proposed system ensures feedbacks to the Arabic learners utilizing deep learning-based models in the level of word and sentence. Experiment results show that the best model performs nearly 4{\%} of error rate for phoneme detection, and roughly 70{\%} of F1-score for mispronunciation detection. \citep{yang2022improving} propose improved mispronunciation detection system based on online pseudo-labeling and wav2vec model for English. In details, pseudo-labeling procedure is performed using unlabeled L2 speech and pre-trained self-supervised learning model (wav2vec) is carried out for enhancing fine-tuning approach. Experiment results demonstrate remarkable score when compared to traditional offline pseudo-labeling techniques. The model exhibits reduction in phoneme error rate nearly 5{\%} and approximately 2.5{\%} enhancement in F1-score for detecting mispronunciation of L2 speech. \citep{peng2022text} present a text aware model for detection of mispronunciation by enhancing weight of audio features compared to score of unrelated text features. For this purpose, contrastive learning procedure is carried out in TIMIT and L2-Arctic English data sets. They report that proposed model ensures roughly 4{\%} improvement comparing with the baselines. 

Our study differs from aforementioned literature works as well mainly because we concentrate on the ensemble-based strategy to enable better accuracy with low computational load compared to high-performance deep learning models. Furthermore, our own dataset of Arabic letters is composed for the purpose of accelerating future studies.

\section{Proposed Framework}\label{sec3}

\begin{figure}[htb!]
\begin{center}
\includegraphics[width=0.45\linewidth]{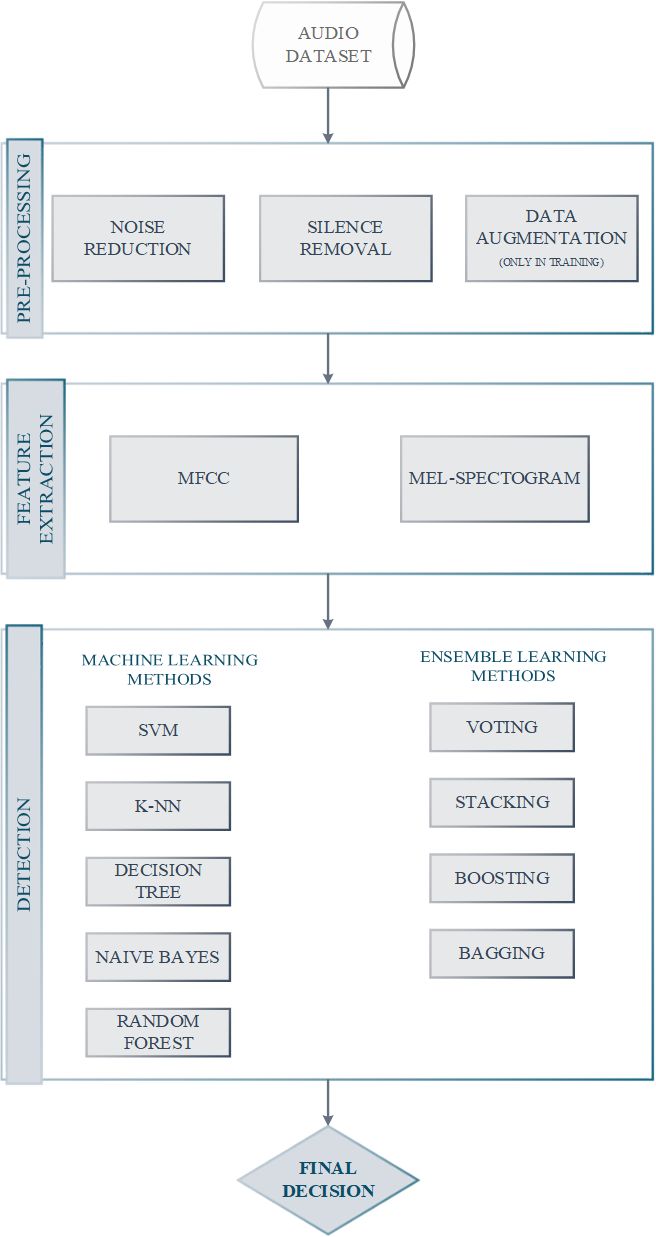}
\caption{The flowchart of the proposed model.}
\label{Figure1}
\end{center}
\end{figure}

\begin{figure}[h!]
\begin{center}
\includegraphics[width=0.55\linewidth]{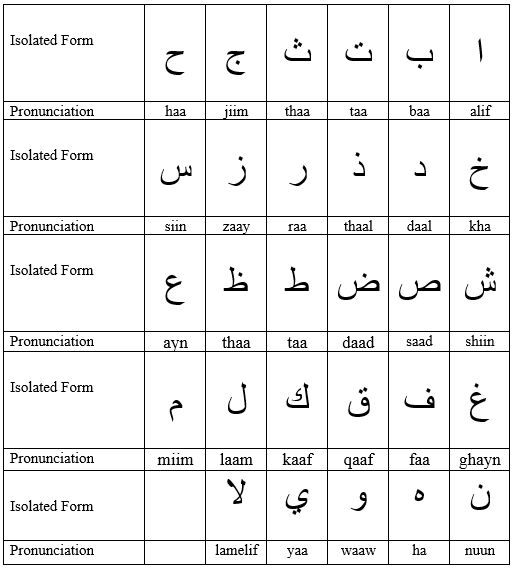}
\caption{Isolated form and pronunciation of Arabic letters.}
\label{Arabic Letters}
\end{center}
\end{figure}

In this work, an ensemble machine learning based mispronunciation detection method for Arabic phonemes is proposed. Flowchart of the proposed system is depicted in Figure \ref{Figure1}. The proposed system has three main stages. First step is to utilize preprocessing to improve the sound quality and the performance of the next steps. Next, feature extraction process is applied to the raw audio signals obtained from the microphone. Finally, the performance of machine learning-based approaches on the problem is examined. Then, the performance is improved by using the ensemble learning approach on these methods. Arabic phonemes used in this work is Arabic letters which is given in Figure \ref{Arabic Letters}. In this figure isolated forms and pronunciation of letters are presented.

\subsection{Pre-processing}
\begin{figure}[h!]
\begin{center}

    \begin{subfigure}{0.5\textwidth}
    \includegraphics[width=\linewidth]{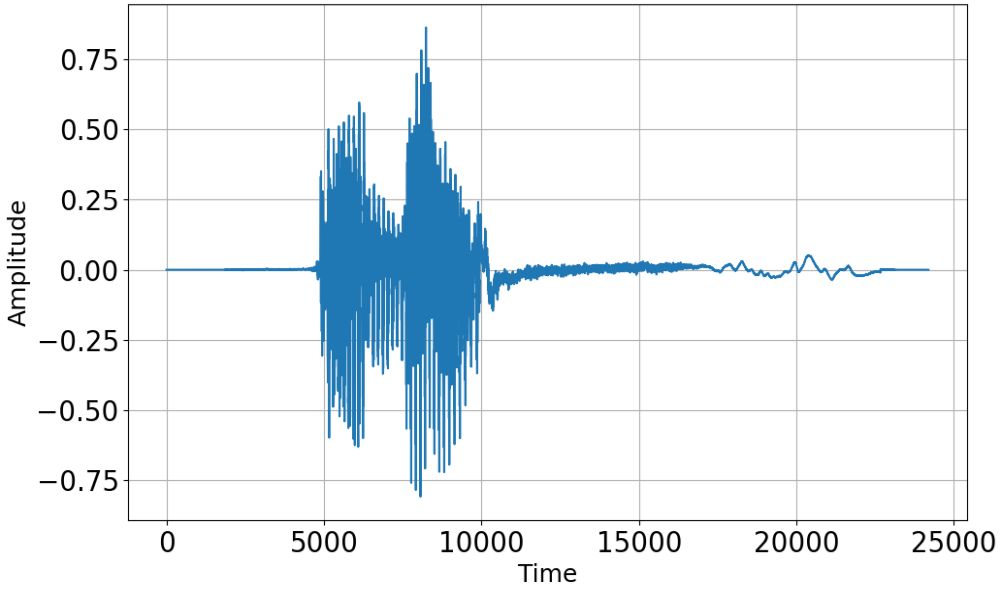}
    \caption{}
    \end{subfigure}\hfil 
    \begin{subfigure}{0.5\textwidth}
    \includegraphics[width=\linewidth]{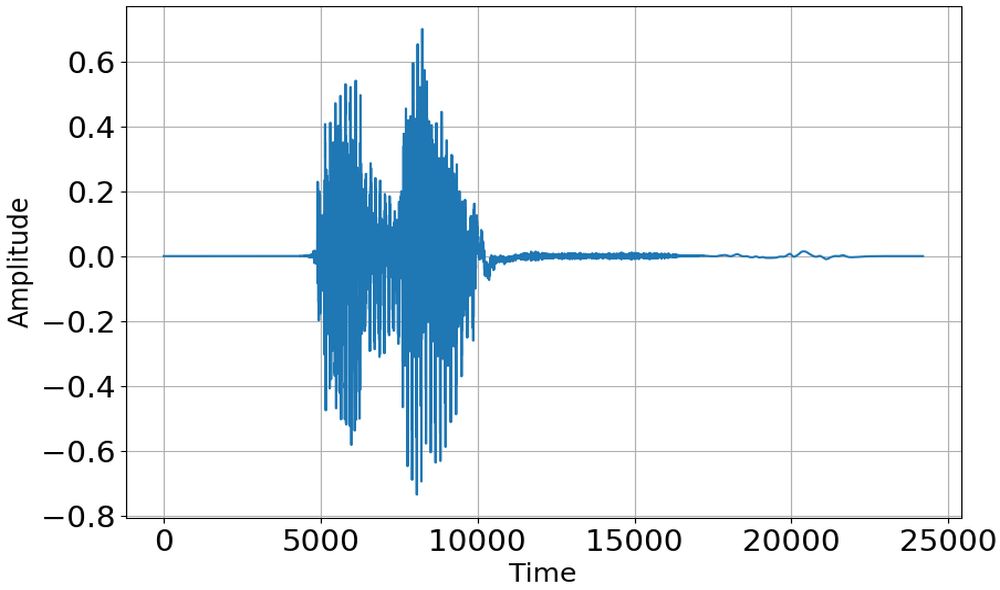}
    \caption{}
    \end{subfigure}\hfil 
    \begin{subfigure}{0.5\textwidth}
    \includegraphics[width=\linewidth]{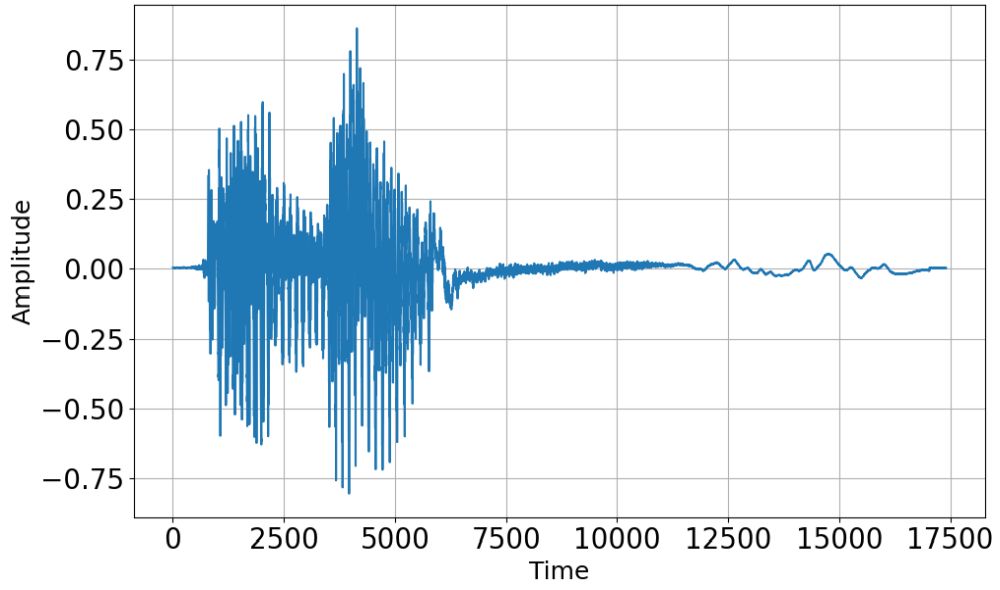}
    \caption{}
    \end{subfigure}\hfil 
    \begin{subfigure}{0.5\textwidth}
    \includegraphics[width=\linewidth]{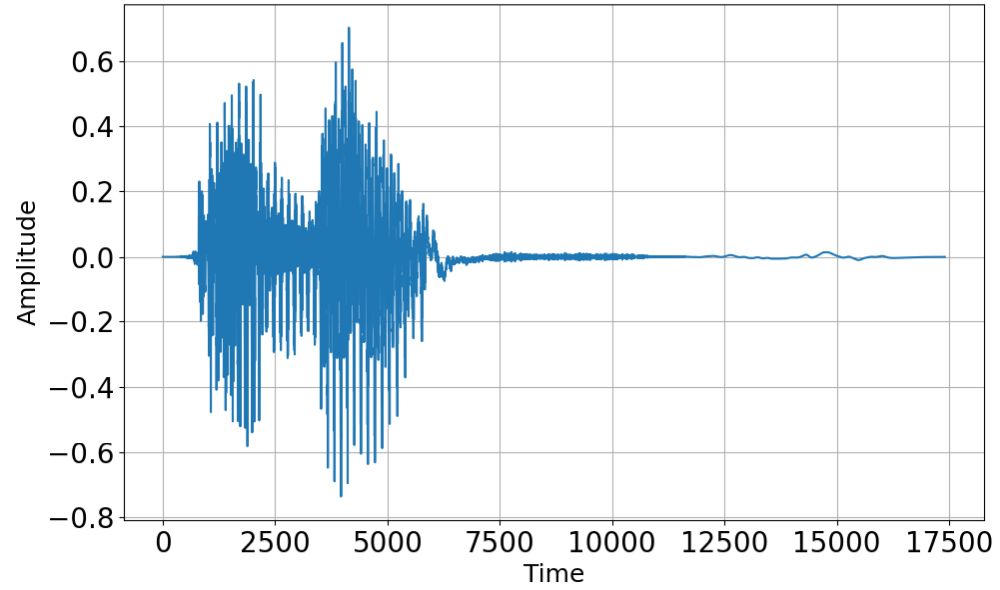}
    \caption{}
    \end{subfigure}
\caption{Noise reduction and silence removal pre-processing operations (a) Original audio signal (b) After noise reduction (c) After noise silence removal (d) After noise reduction and silence removal.}
\label{Pre-processing}
\end{center}
\end{figure}

The pre-processing step aims to separating the clean speech from a mixed sound of speech and noise and get trimming only the part with the audio signal. Noise reduction which relies on a method called “spectral gating” which is a form of Noise Gate is primarily used to improve speech recognition. Then, silence removal process is applied. The beginning of speech indicates that the sound level rises above a certain dB value. In this study, the segment where the audio signal first rises above 30 dB and then drops below 30 dB is trimmed and used as an input to the algorithms. Figure \ref{Pre-processing} shows a sample input signal before and after noise reduction and silence removal operations. As the last preprocessing operation, the number of data is increased by augmentation process. Noise adding, time shifting, time stretching and pitch shifting are the used augmentation techniques. This process is not included in the testing process, but only in the training process.

\subsection{Feature extraction}
Human speech has various distinctive features. However, processing raw audio signal is generally not preferred due to the high noise and data size. It is observed that extracting distinctive features from the audio signal and using it as input to the base machine learning model will produce much better performance than directly considering raw audio signal as input. MFCC \citep{Dave2013} and Mel-Spectrogram \citep{Stevens1937} are two widely used technique for extracting the features from the audio signal. 

The Mel-spectrogram consist of 128 Mel feature and Fast Fourier Transform (FFT) with 2048 window length. The dimensions of the obtained features differ according to the audio duration. Therefore, the feature vector size is scaled to a single dimension. After the Mel features are obtained, it is scaled to 1×128 by taking the averages (before scaling the dimensions are like 128×48, 128×40 for example). In the end, the problem evolves into the classification of the feature vectors obtained in size 1×128.

MFCC features are the result of a series of applied processes. These processes are as follows: Windowing of the audio signal, applying Discrete Fourier Transform (DFT), obtaining the receipt of the logarithm of amplitude, distorting frequencies on a Mel scale, and applying inverse discrete cosine transform (DCT). As a result of these processes, 39 features are obtained, of which the last 20 have more distinctiveness. In this study, these last 20 features are used as input features. Similar to the Mel-spectrogram, the audio file size causes a change in the dimensions of the features. For this reason, all features are averaged among themselves and the problem turns into the classification of the data in the size of 1×20. 

\begin{figure}[h!]
\begin{center}

    \begin{subfigure}{0.41\textwidth}
    \includegraphics[width=\linewidth]{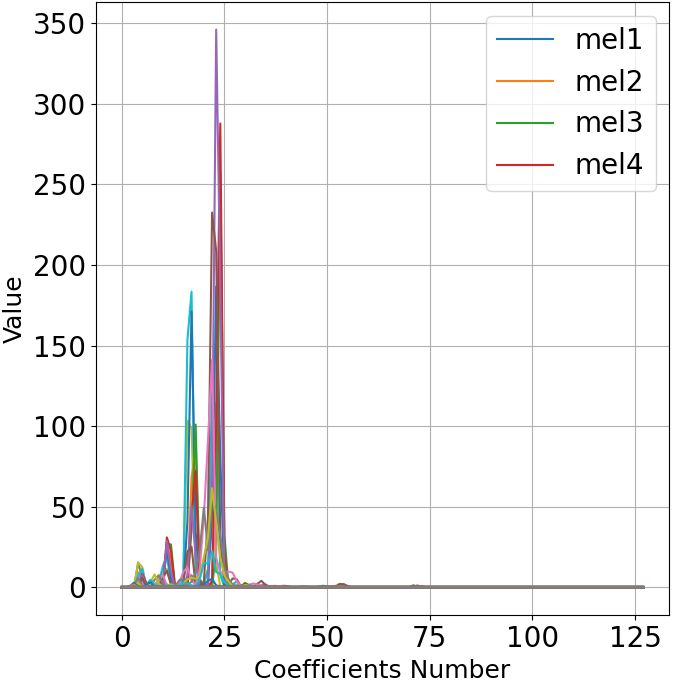}
    \caption{}
    \end{subfigure}\hfil 
    \begin{subfigure}{0.41\textwidth}
    \includegraphics[width=\linewidth]{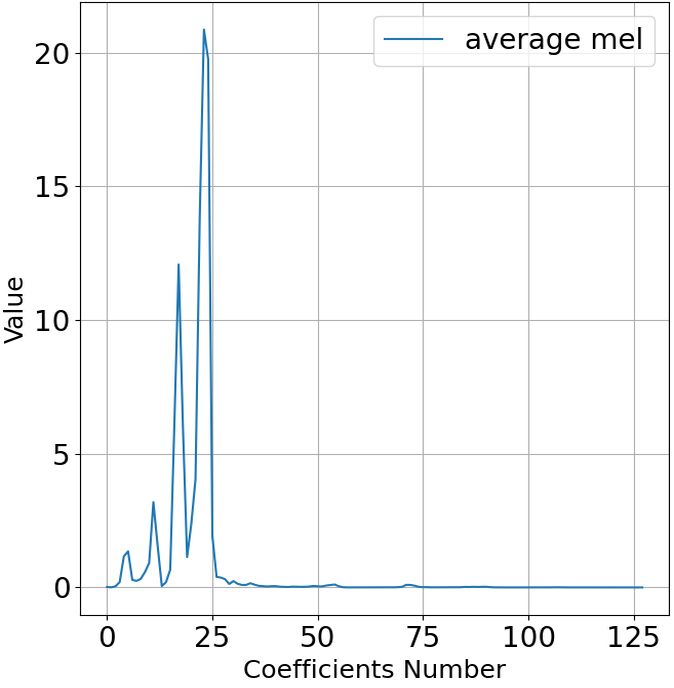}
    \caption{}
    \end{subfigure}\hfil 
    \begin{subfigure}{0.41\textwidth}
    \includegraphics[width=\linewidth]{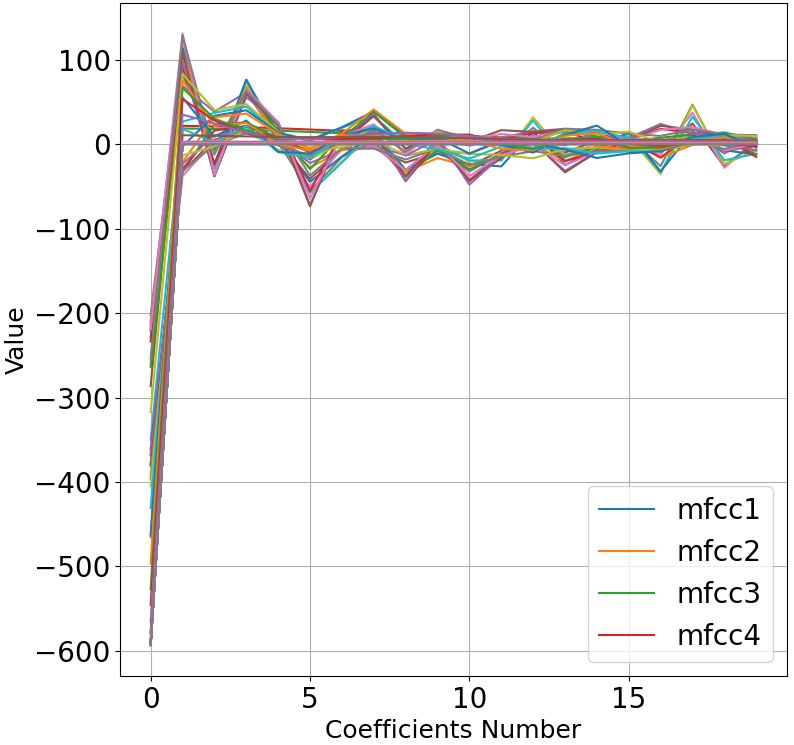}
    \caption{}
    \end{subfigure}\hfil 
    \begin{subfigure}{0.41\textwidth}
    \includegraphics[width=\linewidth]{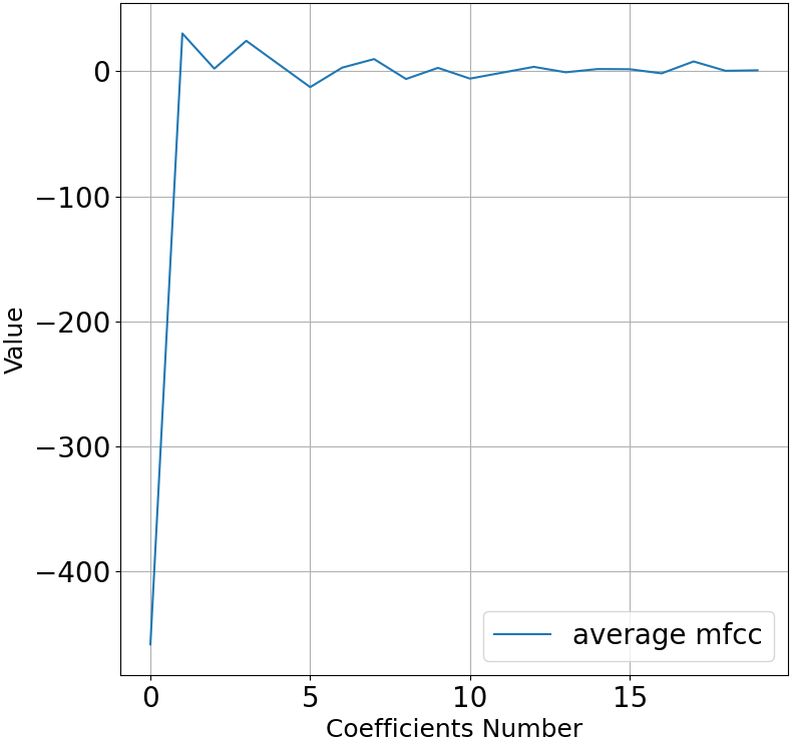}
    \caption{}
    \end{subfigure}
\caption{Scaling of feature vectors (a) Original Mel-spectrogram image (b) Averaged Mel-spectrogram image (c) Original MFCC image (d) Averaged MFCC image.}
\label{MFCC and Mel Features}
\end{center}
\end{figure}

\begin{figure}[h!]
\begin{center}

    \begin{subfigure}{0.5\textwidth}
    \includegraphics[width=\linewidth]{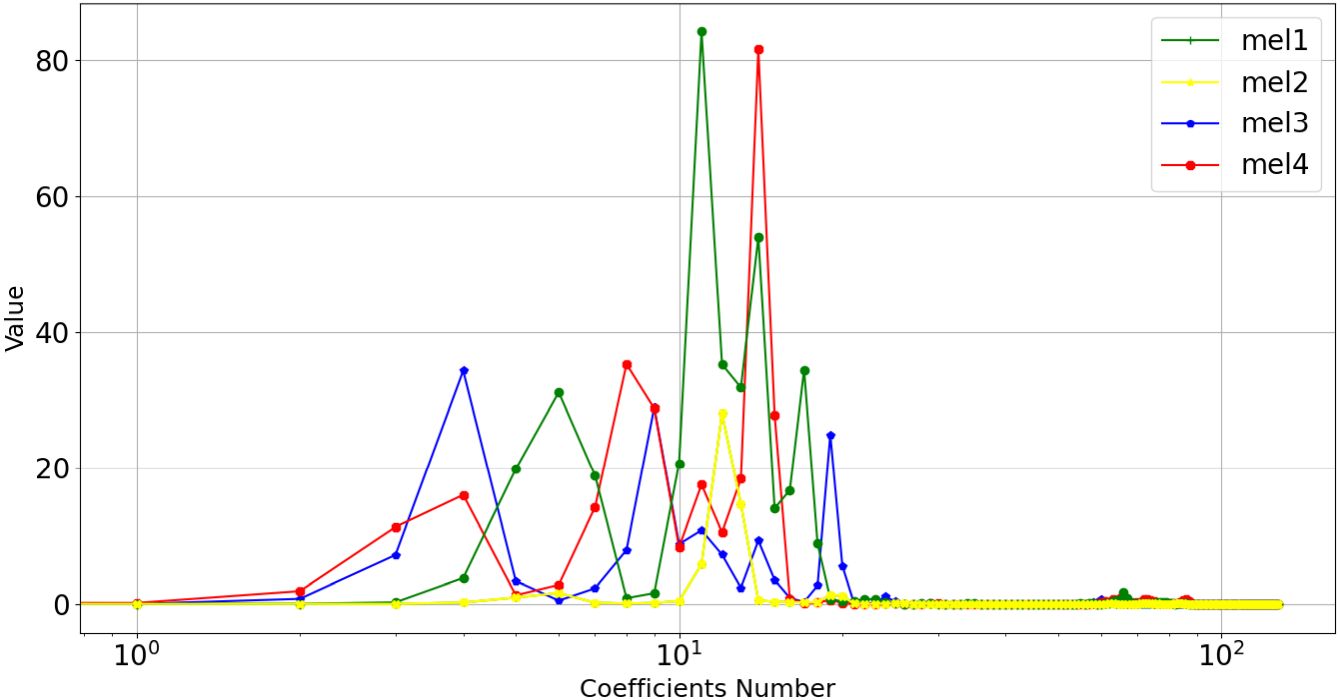}
    \caption{}
    \end{subfigure}\hfil 
    \begin{subfigure}{0.5\textwidth}
    \includegraphics[width=\linewidth]{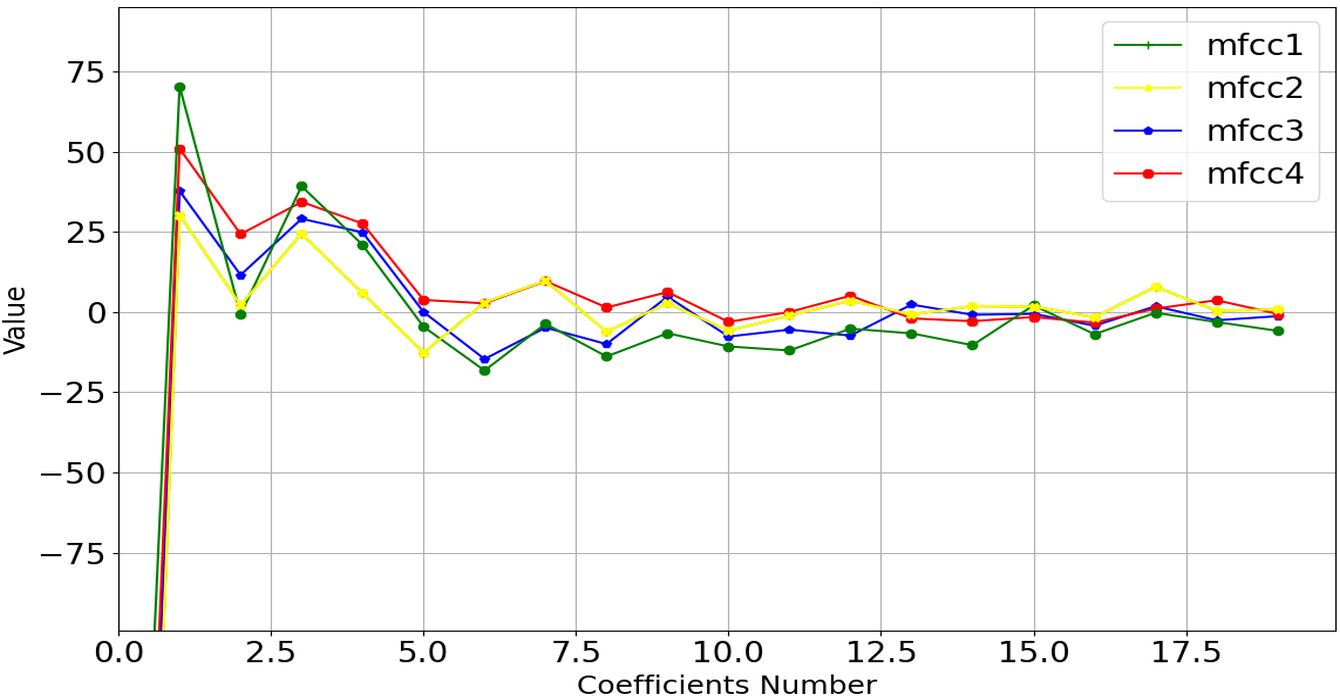}
    \caption{}
    \end{subfigure}\hfil 
    \begin{subfigure}{0.5\textwidth}
    \includegraphics[width=\linewidth]{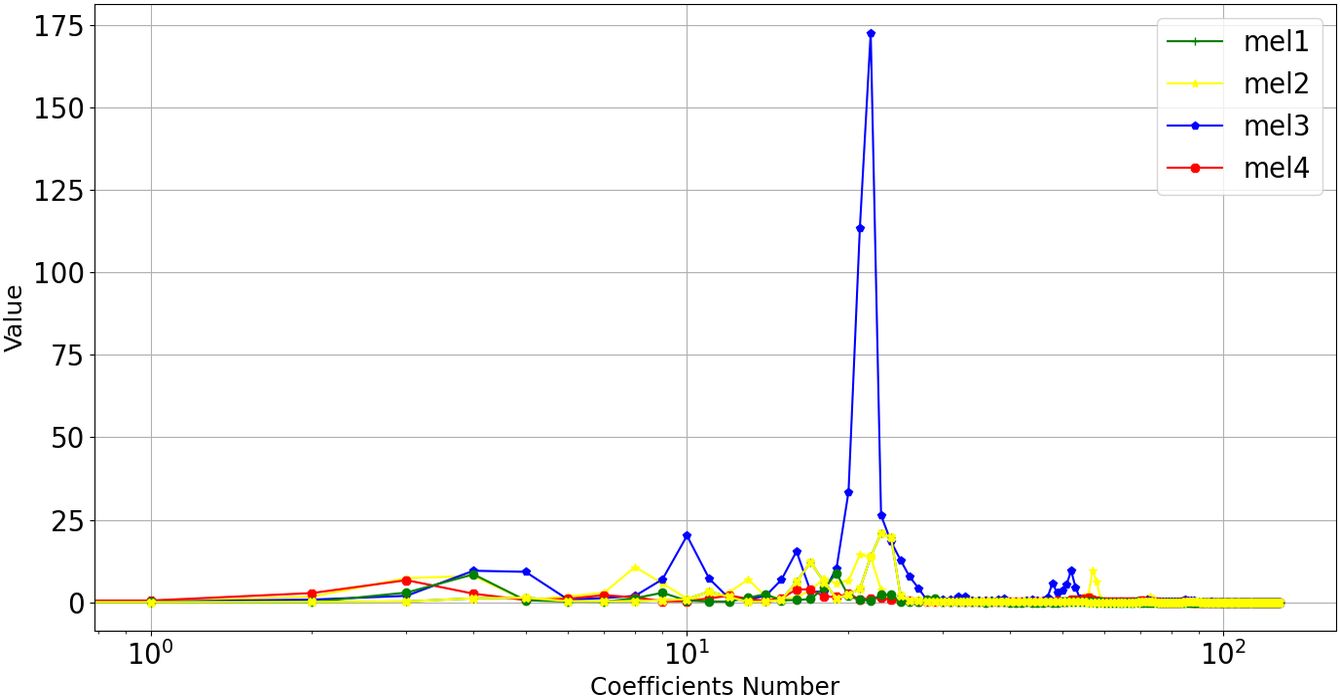}
    \caption{}
    \end{subfigure}\hfil 
    \begin{subfigure}{0.5\textwidth}
    \includegraphics[width=\linewidth]{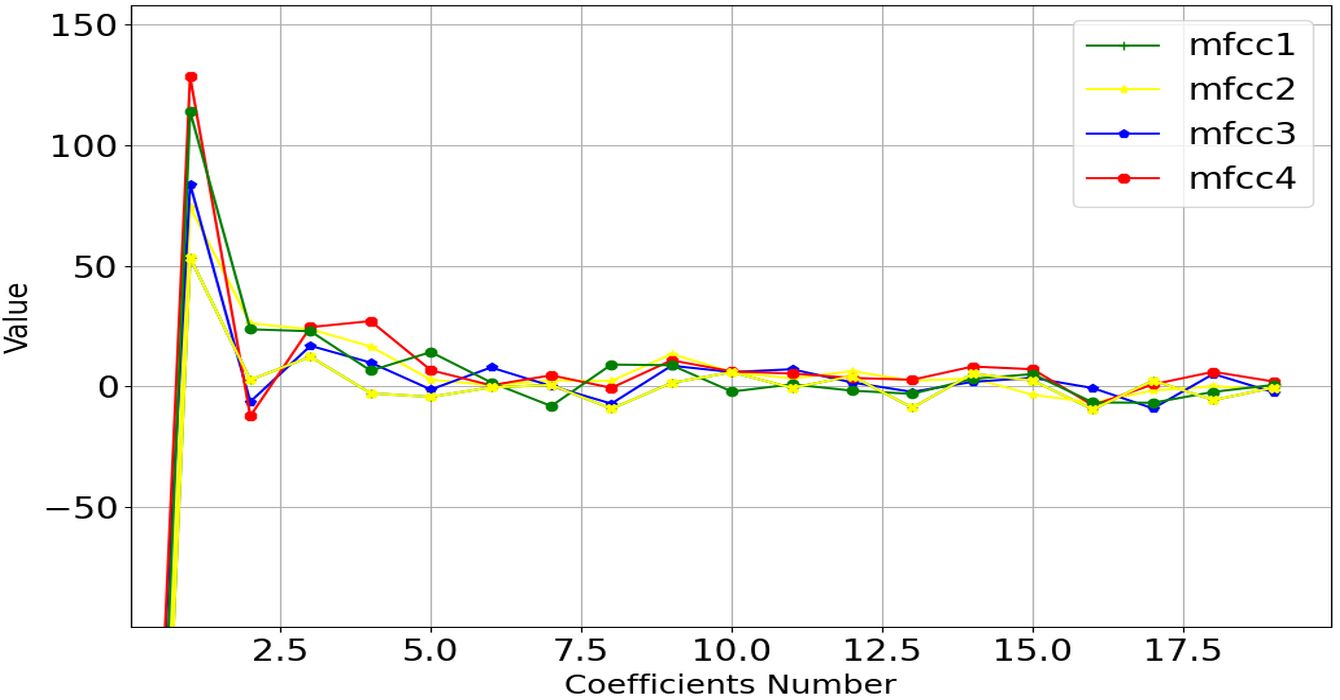}
    \caption{}
    \end{subfigure}
\caption{Average feature vectors of “Alif” and “Baa” letters (a) Average Mel-spectrogram feature of “Alif” (b) Average MFCC feature of “Alif” (c) Average Mel-spectrogram feature of “Baa” (d) Average MFCC feature of “Baa”.}
\label{Alif and Baa Features}
\end{center}
\end{figure}

Figure \ref{MFCC and Mel Features} shows the scaling process of feature vectors and Figure \ref{Alif and Baa Features} presents average Mel-Spectrogram and MFCC features for 2 letters (“Alif” and “Baa”). In Figure \ref{Alif and Baa Features}, different colours represent the speeches of different hafizes. As seen from this figure, even if different hafizes say the same letter, it creates dissimilar characteristic on the formed feature vector. In addition, it is seen that the distinctiveness of the same people in different letters is not sufficient. Because of these difficulties, the performance of the classification method is become more important.

\subsection{Detection}

\subsubsection{Machine learning methods}

Machine learning can be defined as artificial intelligence algorithms that can infer and predict from data to mimic the way humans learn. There are various machine learning algorithms that are capable of solving classification, regression and clustering tasks. In our work, popular methods suitable for classification problem are emphasized which are support vector machine (SVM), k-Nearest neigbour (k-NN), decision tree (DT), naïve Bayes, random forest.

\textbf{SVM} \citep{Cristianini2000} is a supervised learning approach. Kernel functions can also be used depending on the type of data during the operation of the algorithm. In this way, both linear and nonlinear classification operations can be performed. It is aimed to separate all data with a hyperplane. However, if the data cannot be fully separated, they cannot be classified with a single plane. Therefore, different kernel functions are used. A margin is determined around the hyperplane. Whether this margin is large or small directly affects the classification performance. Margin can be controlled with the “C” hyperparameter. The larger the C, the narrower the margin. Also, if the model is overfit, C needs to be reduced. In this work linear kernel and 0.02 used as C parameter. 

\textbf{k-NN} \citep{Mucherino2019} is basically based on the determination of the class of the data whose class is unknown, according to the nearest “k” neighbor from the data in the training set. As a result of performing a distance measurement between the test data and the training data, the nearest “k” nearest neighbors are determined. Then, the class value of the tested data is determined according to these labels. In this work, 3,4 and 5 is evaluated as “k” value.

Basic purpose of \textbf {decision trees} \citep{Alp2019Makine} is to divide the data set into smaller subgroups that are more visually understandable within the framework of certain rules (decision rules). Since the output of the algorithm is a flowchart that looks like a tree visually, it is called a decision tree. There are 4 basic structures on a decision tree: root node, nodes, branches and leaves (terminal node). The root node is where classification process starts from this point. If the observations are in a homogeneous structure, they will naturally be in the same class and the classification process will end without branching the root node. In heterogeneous observations, the root node divides into two or more branches according to the best quality that divides the observations into classes and creates new nodes. The last non-branching node of the tree is the terminal node and represents the classes to which the observations are assigned.

\textbf {Naïve Bayes} \citep{Webb2010} classification is based on Bayes theorem. It is used to estimate the probability that a particular set of features belongs to a particular class. It aims to select the decision with the highest probability using probability calculations. Each attribute is considered independent from other attributes in the class. different class based on various attributes. Naïve Bayes classifiers are extremely fast compared to more complex methods.

\subsubsection{Ensemble learning techniques}
Ensemble learning is an approach to boost the overall accuracy of a classification framework by utilizing multiple learning algorithms. The strategy usually yields better results than a individual learning model. There are most commonly employed ensemble learning techniques in the literature namely, bagging, stacking, boosting, and voting. 

\textbf {Bagging} is one of the most widely-applied ensemble based algorithms \citep{Breiman1996}. It is known as the abbreviation of bootstrap aggregating. In this approach, diversity is performed by re-sampling in which various training data subsets are randomly chosen with replacement from the whole training data set. Each subset is assessed to train a distinct base learner in the set of ensemble learners. Final decision is determined by combining decisions of individual learners by taking a majority vote.

\textbf {Boosting} \citep{Freund1996} is another ensemble learning model that is asserted as an alternative to the bagging technique. The main approach behind of this approach is to produce a set of individual classifiers that utilizes a data subset in which each instance is consolidated with a weight. It is carried out iteratively running base learners on different distributions over the training data set. While all instances have same weight at the first step, the weights of miscategorized instances are updated at each iteration according as the training error of preceding base learners. Each learner employs a subset of instances acquired from an updated version of training data set. After that, instances that are incorrectly forecasted by preceding learners are picked up more frequently than the instances that are correctly forecasted. The final outcome is achieved by weighted majority voting of the categories forecasted by the base learner. AdaBoost, AdaBoost.M1, AdaBoost.M2, AdaBoost.R, Arcing and Real Adaboost \citep{Rokach2010}, \citep{Polikar2006}, \citep{Gopika2014}, \citep{Ren2016} are the variations employed in the literature. AdaBoost.M1 is used in this work.

\textbf {Stacking} is the process of fitting multiple types of models to the same data and then employing metal level model to learn how to integrate the predictions in the best way possible \citep{Dzeroski2004}. There are two major steps. The first one is consists of generating a set of individual classifiers utilizing training set. Thus, meta-data set is composed of the decisions of individual classifiers. After construction of the meta-data set, meta-level learner modeled with this data set in order to get generalized estimations. In our work, the meta-data set comprises of both original training examples and decisions of individual classifiers. Moreover, stacking method is employed with different learning algorithms namely, logistic regression, random forest, and extra tree in our study.

In \textbf {majority voting}, each model makes a prediction (vote) for each test sample, and the prediction of the final result is the model that receives the most votes.

\textbf{Random forest} \citep{Breiman2001} is a set of decision tree classifiers. It is a certain enforcement of bagging method in which each individual classifier is a decision tree. Bagging is utilized to choose training sub sets for each base decision tree. The division task employed in Random forests varies from well-known decision tree strategy where each node is divided by the best attribute among all other attributes. In Random forest approach, a random subset of attributes is chosen first and then the best division is determined on the random subset of attributes. This approach ensures an extra randomness to the model in addition to bagging. Random forests is able to cope with over-fitting problem due to randomness performed in both sample and attribute spaces.

\begin{figure}[h!]
\begin{center}

    \subcaptionbox{}%
    [.47\linewidth]{\includegraphics[width=\linewidth]{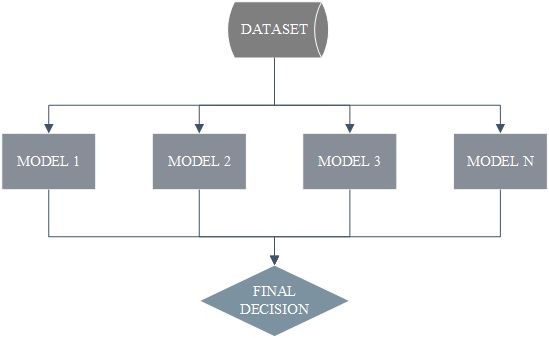}}
    \subcaptionbox{}%
    [.47\linewidth]{\includegraphics[width=\linewidth]{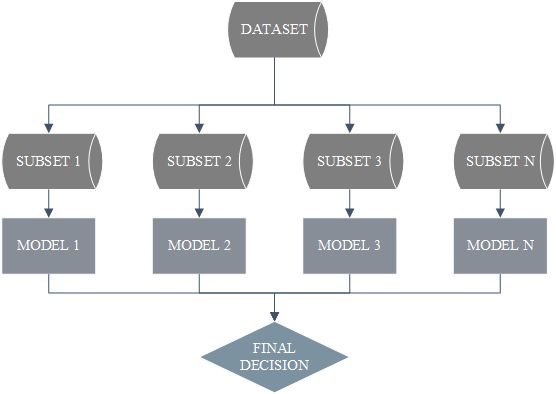}}
    \subcaptionbox{}%
    [.47\linewidth]{\includegraphics[width=\linewidth]{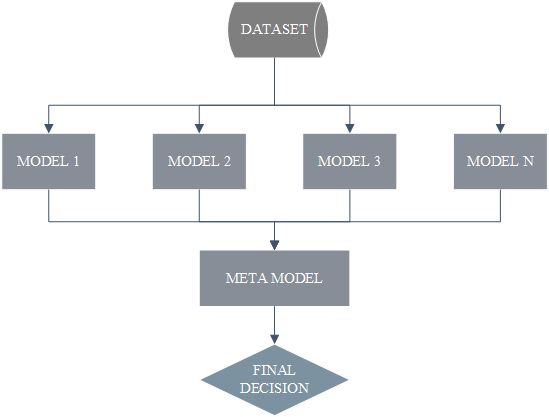}}
    \subcaptionbox{}%
    [.47\linewidth]{\includegraphics[width=\linewidth]{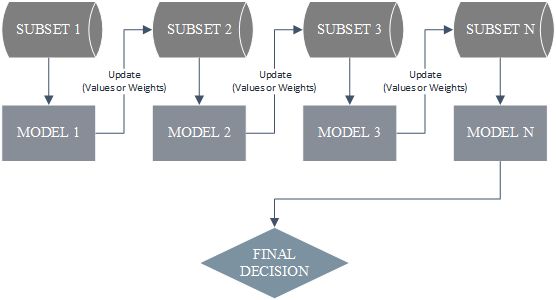}}
\caption{Ensemble architectures (a) Voting (b) Bagging (c) Stacking (d) Boosting.}
\label{Ensemble}
\end{center}
\end{figure}

Figure \ref{Ensemble} shows an overview of the ensemble architectures.

\section{Experimental results}\label{sec4}

In this study, an original dataset is obtained by collecting audio samples from 11 speakers, 8 of whom are hafiz. Dataset consists of Arabic letters. All letters are recorded by asking the speakers to say each letter individually. Original dataset size is 290. Sample size in the dataset is increased to 1450 by using noise adding, time shifting, time stretching, pitch shifting augmentation methods. The dataset is divided into 80{\%} training and 20{\%} test. 5-fold cross-validation procedure is used to ensure that the score of proposed models does not depend on the way we select train and test subsets. The performance of the proposed methods is measured using the formulas given in (1), (2), (3) and (4). In these equations, if we named every class as $ C_{x} $, TP (True Positive) represents audio belonging to the $ C_{x} $ is correctly classified as $ C_{x} $. FP (False Positive) is all non-$ C_{x} $ samples classified as $ C_{x} $. TN (True Negative) denotes all non- $ C_{x} $ samples not classified as $ C_{x} $. FN (False Negative) represents all $ C_{x} $ samples not classified as $ C_{x} $. 

\label{equations}
\begin{equation}
Accuracy = \frac{TP+TN}{TP+TN+FP+FN}
\end{equation}

\begin{equation}
Precision = \frac{TP}{TP+FP}
\end{equation}

\begin{equation}
Recall = \frac{TP}{TP+FN}
\end{equation}

\begin{equation}
F-measure = 2 \times \frac{Precision \times Recall}{Precision + Recall}
\end{equation}

Evaluation results of machine learning based methods are given in Table 1 and Table 2. Table 1 shows effect of MFCC features used as input of machine learning methods. Likewise, Table 2 shows effect of Mel-Spectrogram features as input. In these tables, values with bold font represent best results. According to Table 1 SVM has the best results among the other methods. In Table 2, it can be considered that k-NN where k is determined as 3 is successful. When Table 1 and Table 2 are evaluated together, the highest result is obtained with the k-NN method, in which Mel-Spectrogram features are used as input features.

\begin{table}[hbt!]
\caption{Evaluation results of machine learning methods with MFCC features}
\centering
\begin{tabular}{ccccc}
\hline
\multicolumn{1}{c}{\textbf{Method}} & \multicolumn{1}{c}{\textbf{Accuracy}} & \multicolumn{1}{c}{\textbf{Precision}} & \multicolumn{1}{c}{\textbf{Recall}} & \multicolumn{1}{c}{\textbf{F1-Score}} \\ \hline
SVM                                 & \textbf{0.758}                        & \textbf{0.829}                         & \textbf{0.785}                      & \textbf{0.806}                        \\
k-NN (k=3)                          & 0.466                                 & 0.527                                  & 0.466                               & 0.494                                 \\
k-NN (k=4)                          & 0.308                                 & 0.310                                  & 0.308                               & 0.309                                 \\
k-NN (k=5)                          & 0.307                                 & 0.34                                   & 0.307                               & 0.322                                 \\
Decision Tree                       & 0.714                                 & 0.735                                  & 0.710                               & 0.722                                 \\
Naïve Bayes                         & 0.392                                 & 0.445                                  & 0.392                               & 0.417                                 \\
Random Forest                       & 0.756                                 & 0.78                                   & 0.754                               & 0.766                                 \\ \hline
\end{tabular}
\end{table}

\begin{table}[hbt!]
\caption{Evaluation results of machine learning methods with Mel-Spectrogram features}
\centering
\begin{tabular}{ccccc}
\hline
\multicolumn{1}{c}{\textbf{Method}} & \multicolumn{1}{c}{\textbf{Accuracy}} & \multicolumn{1}{c}{\textbf{Precision}} & \multicolumn{1}{c}{\textbf{Recall}} & \multicolumn{1}{c}{\textbf{F1-Score}} \\ \hline
SVM                                 & 0.952                                 & \textbf{0.970}                         & 0.952                               & 0.960                                 \\
k-NN (k=3)                          & \textbf{0.953}                        & 0.968                                  & \textbf{0.953}                      & \textbf{0.960}                        \\
k-NN (k=4)                          & 0.698                                 & 0.734                                  & 0.698                               & 0.716                                 \\
k-NN (k=5)                          & 0.670                                 & 0.700                                  & 0.67                                & 0.684                                 \\
Decision Tree                       & 0.827                                 & 0.852                                  & 0.823                               & 0.837                                 \\
Naïve Bayes                         & 0.538                                 & 0.68                                   & 0.538                               & 0.600                                 \\
Random Forest                       & 0.938                                 & 0.947                                  & 0.938                               & 0.942                                 \\ \hline
\end{tabular}
\end{table}

\begin{table}
\caption{Evaluation results of ensemble learning methods with MFCC features}
\centering
\begin{tabular}{ccccc}
\hline
\textbf{Method}                & \textbf{Accuracy} & \textbf{Precision} & \textbf{Recall} & \textbf{F1-Score} \\ \hline
Voting                         & \textbf{0.772}    & 0.794              & 0.764           & 0.778             \\
Stacking (Logistic Regression) & 0.700             & 0.720              & 0.704           & 0.711             \\
Stacking (Random Forest)       & 0.761             & \textbf{0.804}     & \textbf{0.768}  & \textbf{0.785}    \\
Stacking (Extra Tree)          & 0.731             & 0.767              & 0.747           & 0.756             \\
Boosting                       & 0.744             & 0.793              & 0.740           & 0.765             \\
Bagging                        & 0.750             & 0.781              & 0.762           & 0.771             \\ \hline
\end{tabular}
\end{table}

\begin{table}
\caption{Evaluation results of ensemble learning methods with Mel-Spectrogram features}
\centering
\begin{tabular}{ccccc}
\hline
\textbf{Method}                & \textbf{Accuracy} & \textbf{Precision} & \textbf{Recall} & \textbf{F1-Score} \\ \hline
Voting                         & \textbf{0.959}    & \textbf{0.969}     & \textbf{0.957}  & \textbf{0.962}    \\
Stacking (Logistic Regression) & 0.866             & 0.893              & 0.862           & 0.877             \\
Stacking (Random Forest)       & 0.951             & 0.960              & 0.947           & 0.953             \\
Stacking (Extra Tree)          & 0.938             & 0.944              & 0.939           & 0.941             \\
Boosting                       & 0.871             & 0.932              & 0.855           & 0.891             \\
Bagging                        & 0.943             & 0.953              & 0.935           & 0.943             \\ \hline
\end{tabular}
\end{table}

\begin{table}[h!]
\caption{Performance evaluation with deep learning based methods}
\centering
\begin{tabular}{cccccc}
\hline
\textbf{Method} & \textbf{Accuracy} & \textbf{Precision} & \textbf{Recall} & \textbf{F1-Score} & \textbf{Time (s)}\\ \hline
\citep{Abdoli2019}              & 0.910             & 0.920              & 0.910           & 0.915   & 0.470           \\
\citep{Choi2018}              & 0.900             & 0.900              & 0.900           & 0.900   & 0.612           \\
Proposed Method & \textbf{0.959}    & \textbf{0.969}     & \textbf{0.957}  & \textbf{0.962}  & \textbf{0.091}  \\ \hline
\end{tabular}
\end{table}

\begin{figure}
\begin{center}
\includegraphics[width=0.49\linewidth]{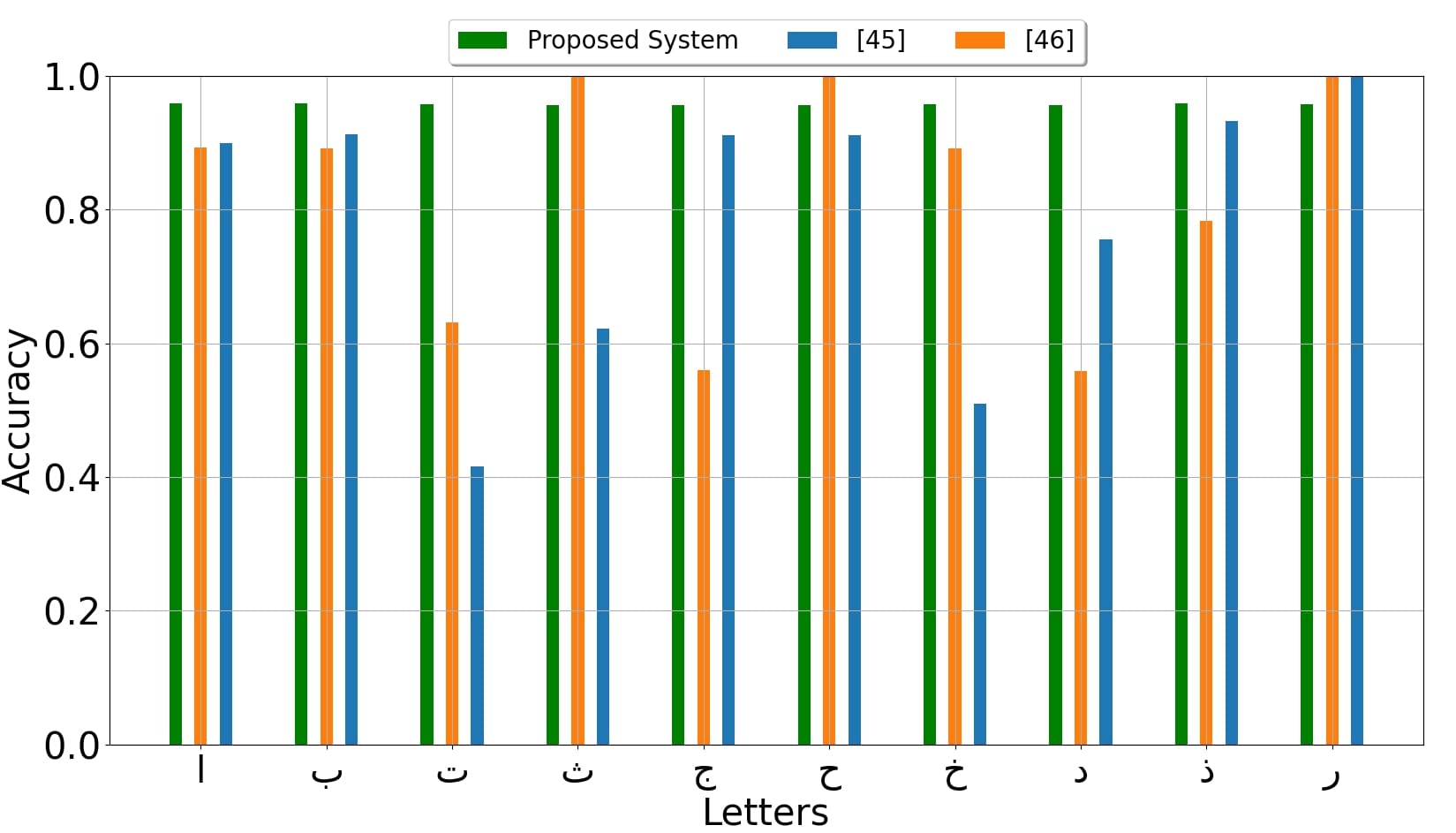}
\caption{Letter-based performance evaluation with deep learning based methods.}
\label{Comparison}
\end{center}
\end{figure}
The ensemble learning approach is applied to the methods given in Table 1 and Table 2. Evaluation results of ensemble learning methods are given in Table 3 and Table 4. In these tables, values with bold font represent best results. When Table 3 is examined, it can be seen that stacking with random forest approach is the best compared to the other approaches. Table 4 clearly shows that voting approach has the best results.

Table 5 shows the comparison of the recent and proposed methods for sound classification in the literature, in terms of processing speed and detection performance. The comparison is carried out on the dataset created within the scope of this study. The methods used in comparison in Table 5 are deep learning-based approaches. These methods are implemented using original network structure and parameters specified in the papers. In this table, proposed method refers voting ensemble method with Mel-Spectrogram features input. As seen in Table 5, proposed method has superior performance results and the lowest processing speed among the other methods. In addition, the letter-based performances of the methods given in Table 5 are analyzed. Figure \ref{Comparison} shows the performances of \citep{Abdoli2019}, \citet{Choi2018} and the proposed method for 10 letters. It is understood that the proposed method shows a steady performance despite the high-performance decrease in some letters in other methods. Table 5 and Figure \ref{Comparison} show that popular deep learning approaches can be surpassed in mispronunciation detection of Arabic phonemes with the proposed ensemble approach. In addition, it should be noted that generally deep learning-based approaches require more training time and processing (inference) time due to their complexity. Arabic mispronunciation detection is a challenging problem. It is significant that the proposed framework ensures superior performance with less computational load to this problem.

\section{Discussion and Conclusion}\label{sec5}

In this work, an ensemble learning approach is proposed to detect mispronunciation of Arabic phonemes. In the proposed method, firstly, feature extraction is performed with MFCC and mel-spectrogram methods. Then, traditional and ensemble learning-based approaches used these features as input and their performance is evaluated on the original data set created for this study. The method with the highest performance obtained from evaluation is also compared with the deep learning-based approaches. Experimental results show that the utilization of voting classifier as an ensemble algorithm with mel-spectrogram feature extraction technique reveals remarkable results with 95.9{\%} of accuracy. Future studies will focus on increasing the data set and transfer learning techniques.


\bibliographystyle{cas-model2-names}

\bibliography{cas-refs}





\end{document}